\documentclass[usenatbib, aps, pra, amsmath, amssymb, 11pt, final, tightenlines, twoside, twocolumn,nofloats, nofootinbib, superscriptaddress, showkeys, showkeywords]
{revtex4}

\usepackage[T2A]{fontenc}
\usepackage[utf8x]{inputenc}
\usepackage[english]{babel}
\usepackage{graphicx}
\usepackage{dcolumn}
\usepackage{bm}
\usepackage{longtable}
\usepackage[usenames]{color}

\def\squareforqed{\hbox{\rlap{$\sqcap$}$\sqcup$}}

\def\sq{\ifmmode\squareforqed\else{\unskip\nobreak\hfil
\penalty50\hskip1em\null\nobreak\hfil\squareforqed
\parfillskip=0pt\finalhyphendemerits=0\endgraf}\fi}

\def\utw{\smash{\rlap{\lower5pt\hbox{$\sim$}}}}

\def\udtw{\smash{\rlap{\lower6pt\hbox{$\approx$}}}}

\def\diameter{{\ifmmode\mathchoice
{\ooalign{\hfil\hbox{$\displaystyle/$}\hfil\crcr
{\hbox{$\displaystyle\mathchar"20D$}}}}
{\ooalign{\hfil\hbox{$\textstyle/$}\hfil\crcr
{\hbox{$\textstyle\mathchar"20D$}}}}
{\ooalign{\hfil\hbox{$\scriptstyle/$}\hfil\crcr
{\hbox{$\scriptstyle\mathchar"20D$}}}}
{\ooalign{\hfil\hbox{$\scriptscriptstyle/$}\hfil\crcr
{\hbox{$\scriptscriptstyle\mathchar"20D$}}}}
\else{\ooalign{\hfil/\hfil\crcr\mathhexbox20D}}%
\fi}}

\newcommand{\pasp}{Publ. Astron. Soc. Pacific }

\input{maik.rty}

\setcitestyle{authoryear,round,aysep={,},citesep={;}}
\begin{document}



\title{THE POST-AGB STAR IRAS\,07253-2001: PULSATIONS, LONG-TERM BRIGHTNESS VARIABILITY AND SPECTRAL PECULIARITIES}

\author{\firstname{N.~P.}~\surname{Ikonnikova}}
\email{ikonnikova@sai.msu.ru} \affiliation{Sternberg Astronomical
Institute, Lomonosov Moscow State University, Moscow, 119234
Russia}

\author{\firstname{M.~A.}~\surname{Burlak}}
\affiliation{Sternberg Astronomical Institute, Lomonosov Moscow
State University, Moscow, 119234 Russia}

\author{\firstname{A.~V.}~\surname{Dodin}}
\affiliation{Sternberg Astronomical Institute, Lomonosov Moscow
State University, Moscow, 119234 Russia}

\author{\firstname{A.~A.}~\surname{Belinski}}
\affiliation{Sternberg Astronomical Institute, Lomonosov Moscow
State University, Moscow, 119234 Russia}

\author{\firstname{A.~M.}~\surname{Tatarnikov}}
\affiliation{Sternberg Astronomical Institute, Lomonosov Moscow
State University, Moscow, 119234 Russia} \affiliation{Faculty of
Physics, Lomonosov Moscow State University, Moscow, 119191 Russia}

\author{\firstname{N.~A.}~\surname{Maslennikova}}
\affiliation{Sternberg Astronomical Institute, Lomonosov Moscow
State University, Moscow, 119234 Russia} \affiliation{Faculty of
Physics, Lomonosov Moscow State University, Moscow, 119191 Russia}

\author{\firstname{S.~G.}~\surname{Zheltoukhov}}
\affiliation{Sternberg Astronomical Institute, Lomonosov Moscow
State University, Moscow, 119234 Russia} \affiliation{Faculty of
Physics, Lomonosov Moscow State University, Moscow, 119191 Russia}

\author{\firstname{K.~E.}~\surname{Atapin}}
\affiliation{Sternberg Astronomical Institute, Lomonosov Moscow
State University, Moscow, 119234 Russia}

\begin{abstract}

The observations and comprehensive study of intermediate initial
mass stars at the late stages of evolution, and after the
asymptotic giant branch (AGB) in particular, are of crucial
importance to identify the common properties for the stars of
given group and to reveal binaries among them. This work aims to
investigate photometric and spectral peculiarities of a poorly
studied post-AGB candidate and infrared source
\mbox{IRAS~07253-2001}. We present the new multicolour
$UBVR_{C}I_{C}YJHK$ photometry obtained with the telescopes of the
Caucasian mountain observatory and analyse it together with the
data acquired by the All Sky Automated Survey for SuperNovae. We
report on the detection of multiperiod brightness variability
caused by pulsations. A beating of close periods, the main one of
73~days and additional ones of 68 and 70~days, leads to amplitude
variations. We have also detected a long-term sine trend in
brightness with a period of nearly 1800~days. We suppose it to be
orbital and IRAS~07253-2001 to be binary. Based on new
low-resolution spectroscopic data obtained with the 2.5-m
telescope of the  Caucasian mountain observatory in 2020 and 2023
in the $\lambda$\,3500--7500 wavelength range we have identified
spectral lines and compiled a spectral atlas. We have found the
[N\,II], [Ni\,II] and [S\,II] forbidden emission lines in the
spectrum and discuss their origin. The H$\alpha$ line has a
variable double-peaked emission component. We have derived
preliminary estimates of the star's parameters and detected a
variation of radial velocity with a peak-to-peak amplitude of
about 30~km\,s$^{-1}$.

\end{abstract}

\maketitle

\section{INTRODUCTION}

One of the most urgent tasks in exploring the evolution of stars
of intermediate initial masses (1--8\,$M_{\odot}$) is to study
these objects in the transition from the asymptotic giant branch
(AGB) to planetary nebulae. During the thermal pulsing of AGB
phase these stars suffer large mass loss and supply the
interstellar medium with nucleosynthesis products created during
the stars' evolution and, thus, along with supernova remnants
stimulate the further development of their host galaxies (Iben and
Renzini, 1983).

As it turned out from observations, the vast majority of objects
in the post-asymptotic (post-AGB) stage of evolution are variable
stars. The type of brightness variability depends on the
temperature of the star, that is, on its position on the
horizontal evolutionary track. Cooler objects pulsate, and, as a
rule, not with one single frequency (Sasselov, 1984; Arkhipova et
al., 2010; Hrivnak et al., 2020) while the hot ones show rapid
(with a characteristic time of several days or less) irregular
variability, which may be due to variations in the stellar wind
power, as well as to the pulsations of the compact core (Handler
et al., 1997; Arkhipova et al., 2013). Besides, the variations of
circumstellar reddening in inhomogeneous dust shells play a
significant role in the occurrence of photometric variability of
post-AGB stars (Arkhipova et al., 2010; Hrivnak et al., 2022).

Currently, the pulsation theory of stars in the late stages of
evolution is under development. It's rather difficult to construct
such a theory because for AGB and post-AGB stars the convection
processes and outflows of matter are of crucial importance, and
these are hard to simulate (Fadeev, 2019).  The pulsation
characteristics obtained from observations for as many stars as
possible provide valuable information for computing pulsation
models.

A considerable portion of the currently known post-AGB objects are
binaries (Van Winckel, 2003, 2007). These stars are not contact
systems at this stage of evolution, but they should have been
subjected to strong interaction in the past, when the main star
was on AGB and had a larger size (Van Winckel, 2017), so it is
important to distinguish between binary and single objects and to
consider them separately when comparing with theoretical
evolutionary models.

Among the 209 most probable post-AGB objects presented in the
catalogue of Szczerba et al. (2007),  there are stars that have
not been investigated well enough. One of them is the infrared
(IR) source IRAS\,07253-2001, which was included in the list of
post-AGB candidates by {Garc\'{\i}a-Lario} et al. (1990).  For the
first time, the authors obtained the $JHK$ observations for the
star and constructed the energy distribution based on these data,
and the IRAS data in the wavelength range from 12 to 60\,$\mu$m as
well. In {Garc\'{\i}a-Lario} et al. (1990), the IR source was
wrongly identified with the bright neighboring star HD\,59049,
nevertheless the $JHK$-observations refer to IRAS\,07253-2001, and
not to HD\,59049.

Later the object was added to the sample of possible OH/IR masers
(Blommaert et al., 1993) but the 1612 MHz emission was not
detected. Neither H$_{2}$O ({Su\'{a}rez} et al., 2007) nor SiO
(Yoon et al., 2014)  maser emission associated with
IRAS\,07253-2001 was found.

Blommaert et al. (1993)  classified IRAS\,07253-2001 as an
oxygen-rich (O-rich) AGB object which was later confirmed by Suh
and Hong (2017) who included the source in the catalogue of O-rich
AGB objects.

Reddy and Parthasarathy (1996)  obtained the $BVI$ photometry and
a low-resolution spectrum for IRAS\,07253-2001. The authors
defined the spectral class as F5\,Ie and derived a satisfactory
model fit to the spectral energy distribution in the wavelength
region from 0.4 to 100\,$\mu$m which was a sum of radiation from
the photosphere with $T_{\rm eff}=7000$~K and $\log g=1$ and the
dust shell heated to $T_{d}=210$~K. At the same time the authors
indicated the presence of both cold and warm dust shells in the
system. According to the calculations of Reddy and Parthasarathy
(1996)  the star has a total $V$ extinction of
$A_{V}=2\,.\!\!^{\rm m}1$, radius of $R_{*}=54R_{\odot}$, is
surrounded by a dust shell with
\mbox{$R_{d}=1.0\times10^{5}R_{\odot}$} and is located at a
distance of $d=10$~kpc.  {Su\'{a}rez} et al. (2006) classified
IRAS\,07253-2001 as an F2 supergiant based on a low-resolution
spectrum.

Our aim was to study the photometric behaviour of the star and its
spectral peculiarities, and to determine the star's parameters
based on both archive and our new data. Here we present the
analysis of photometric and spectroscopic data for the star
obtained with the telescopes of the Caucasian mountain observatory
of the Sternberg astronomical institute of the Lomonosov Moscow
State University (CMO SAI MSU) and the photometry acquired by the
All Sky Automated Survey for SuperNovae (ASAS-SN). We report on
the detection of photometric and spectral variability and estimate
the star's parameters.

The obtained observational data have demonstrated that the
equipment, weather conditions and the skills of the staff at the
CMO SAI MSU have proved suitable to get good quality data for the
observational project devoted to post-AGB stars and related
objects which was started on the telescopes of the Crimean
astronomical station of SAI MSU more than 30 years ago.

\section{OBSERVATIONS}

\subsection{$UBVR_{C}I_{C}$-photometry}

Optical photometry for the star was obtained on the 60-cm
Ritchey-Cr\'{e}tien telescope (RC600) at the CMO SAI MSU. The
telescope is equipped with a set of photometric filters and an
Andor iKon-L CCD (2048$\times$2048 pixels of 13.5~$\mu$m, the
pixel scale is $0\,.\!\!^{\prime\prime}67$~pixel$^{-1}$, the field
of view is $22^\prime\times22^\prime$). For a more detailed
description of the telescope and instrumentation we refer to
Berdnikov et al. (2020). The observations were carried out in
remote control mode. We have observed IRAS\,07253-2001 for four
seasons of visibility in 2019--2023. A complete set of exposures
for each night consisted of 2--3 frames in each of the
$UBVR_{C}I_{C}$ filters.

On one photometric night (January~2, 2020) we made a series of
frames at close airmass for the standard field SA104 presented in
Landolt (2009).  Astrometry and photometry for the standard stars
from the field were taken from the database of Peter
Stetson\footnote{\url{https://www.canfar.net/storage/list/STETSON/
Standards/L104}}. Based on the photometry for the standard field,
we derived the transformation coefficients to calibrate our
instrumental photometry to the standard system. Then we selected a
sample of rather bright stars of \mbox{$V=11\,.\!\!^{\rm
m}0$--$14\,.\!\!^{\rm m}5$} in the vicinity of IRAS\,07253-2001
and transformed their instrumental magnitudes to the standard
system using the derived coefficients. Based on the reduced data
for the IRAS\,07253-2001 field we chose two stars with brightness
and colours close to those of IRAS\,07253-2001 to be used later as
comparison stars in differential photometry. According to the
ASAS-SN database the selected stars did not show variability on a
timescale of about 4000~days. Fig.~1 provides a finding chart for
IRAS\,07253-2001 and comparison stars. Their 2MASS designations
and our derived $UBVR_{C}I_{C}$ magnitudes are listed in Table~1.

\begin{table*}
    \caption{$UBVR_{C}I_{C}$-photometry for the comparison stars}
    \label{tab:tabl1}
    \begin{tabular}{c|c|c|c|c|c|c}
        \hline
        Star & ID 2MASS & $U$ & $B$ & $V$ & $R_{C}$ & $I_{C}$    \\
        \hline
        St99 & 07275650-2007120 &  14.598 & 14.243 & 13.466 & 13.051 & 12.641 \\
        St122 & 07271826-2009324 & 13.494 & 13.419 & 12.797 & 12.443 & 12.083\\
        \hline
    \end{tabular}
\end{table*}

\begin{figure*}
    \centering
    \includegraphics[scale=0.9]{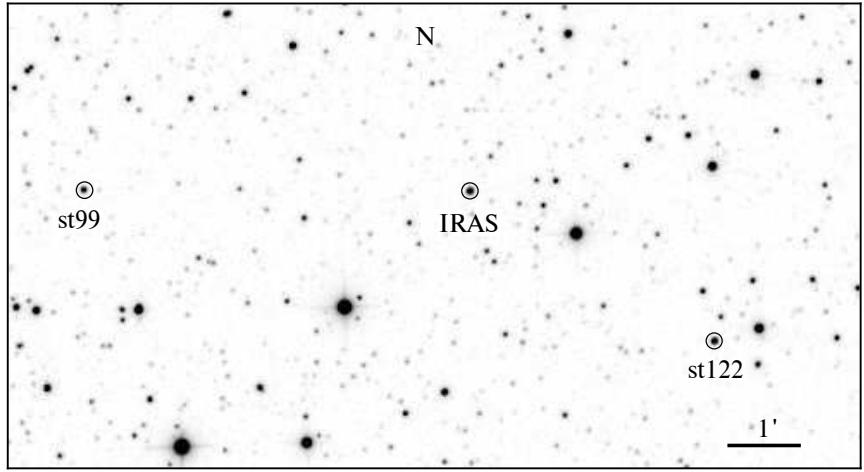}
    \caption{Finding chart for the field of IRAS\,07253-2001 in $V$.}
    \label{fig1}
\end{figure*}

We present the resulting photometry for IRAS\,07253-2001 in
Table~2 (in its entirety the table is provided in electronic form
(\url{http://lnfm1.sai.msu.ru/~davnv/iras07253/UBVRcIc.txt}))
where for every night we list the mean time of observation and
magnitudes in each of photometric bands averaged over 2--3 frames.
Our uncertainties defined as standard deviations for each night
and averaged over all nights are $\Delta U=0\,.\!\!^{\rm m}023$,
$\Delta B=0\,.\!\!^{\rm m}007$, $\Delta V=0\,.\!\!^{\rm m}008$,
$\Delta R_{C}=0\,.\!\!^{\rm m}010$, $\Delta I_{C}=0\,.\!\!^{\rm
m}007$.

\begin{table}[h!]
\centering \caption{$UBVR_{C}I_{C}$-photometry for \mbox{IRAS\,07253-2001} in 2019--2023} \label{tab:tabl2}
\begin{tabular}{c|c|c|c|c|c}
\hline
 JD, 2400000+&$U$&$B$&$V$&$R_{C}$&$I_{C}$\\
 \hline
          58784.610&    13.736& 13.449& 12.675& 12.199& 11.708\\
          58785.617&    13.698& 13.479& 12.676& 12.207& 11.699\\
          58791.602&    13.603& 13.313& 12.569& 12.111& 11.638\\
          ...&        ...&    ...&    ...&    ...&    ...   \\
\hline
\end{tabular}
\end{table}

\subsection{IR-photometry}

Near-IR photometry was carried out on the 2.5-m telescope of the
CMO SAI MSU with the ASTRONIRCAM camera-spectrograph  (Nadjip et
al., 2017) during six seasons of visibility in 2018--2023. We used
the dithering mode to obtain images in the $JHK$ bands of the
\mbox{MKO--NIR} system (Mauna Kea Observatories Near-InfraRed
(Simons and Tokunaga, 2002; Tokunaga et al., 2002)) and also in
the $Y$ band in 2021--2023. We took 10--15 images in each filter
for each pointing. The initial processing of raw images described
in detail in Tatarnikov et al. (2023) included the correction for
non-linearity and bad pixels, dark subtraction, flat-fielding and
background subtraction. Then we performed aperture-based
photometry. Usually we used HD\,59049 (A2\,III/IV) as a comparison
star: it is close to IRAS\,07253-2001 in IR brightness and
appeared in the field of view of the camera
($4\,.\!\!^{\prime}6\times4\,.\!\!^{\prime}6$). Its MKO-NIR
magnitudes ($Y=9\,.\!\!^{\rm m}53$, $J=9\,.\!\!^{\rm m}40$,
$H=9\,.\!\!^{\rm m}29$, $K=9\,.\!\!^{\rm m}28$) were calculated
from the 2MASS magnitudes according to the transforming equations
given in Leggett et al. (2006). Sometimes during the 2021--2022
season HD\,59049 was not caught by the camera. Then we used
HD\,59095 (A3\,IV/V) as a comparison star. Its magnitudes
\mbox{($Y=8\,.\!\!^{\rm m}52$}, $J=8\,.\!\!^{\rm m}42$,
$H=8\,.\!\!^{\rm m}37$, $K=8\,.\!\!^{\rm m}37$) were derived
similarly but adjusted so that the brightness differences with
HD\,~59049 corresponded to those observed when both comparison
stars came into view. Tables~3 and 4 present the resulting
$YJHK$-photometry. The magnitudes were calculated as mean values
for each pointing. The errors were computed as standard
deviations, they do not include the uncertainties of the
comparison stars' magnitudes. Their mean values are $\Delta
Y=0\,.\!\!^{\rm m}009$, $\Delta J=0\,.\!\!^{\rm m}014$, $\Delta
H=0\,.\!\!^{\rm m}015$, $\Delta K=0\,.\!\!^{\rm m}013$.


\renewcommand{\baselinestretch}{0.85}
\begin{table}[h!]
    \centering
    \caption{$JHK$-photometry for IRAS\,07253-2001 in 2018--2021}
    \label{tab:tabl3}
    \begin{tabular}{c|c|c|c}
        \hline
        JD, 2400000+ & $J$, mag & $H$, mag & $K$, mag \\
        \hline
       58147.331 & 10.867  & 10.048  & 8.808  \\
        58151.340 & 10.893  & 10.079  & 8.852  \\
        58153.360 & 10.893  & 10.084  & 8.850  \\
        58156.296 & 10.911  & 10.102  & 8.899  \\
        58166.312 & 10.888  & 10.054  & 8.840  \\
        58180.248 & 10.892  & 10.077  & 8.834  \\
        58482.412 & 10.906  & 10.097  & 8.877  \\
        58486.395 & 10.901  & 10.097  & 8.872  \\
        58487.375 & 10.896  & 10.089  & 8.858  \\
        58489.350 & 10.885  & 10.097  & 8.848  \\
        58493.401 & 10.903  & 10.107  & 8.895  \\
        58504.439 & 10.848  & 10.079  & 8.905  \\
        58511.479 & 10.844  & 10.079  & 8.862  \\
        58863.360 & 10.901  & 10.105  & 8.870  \\
        58866.356 & 10.923  & 10.126  & 8.918  \\
        58867.410 & 10.912  & 10.126  & 8.923  \\
        58868.354 & 10.913  & 10.119  & 8.919  \\
        58869.371 & 10.899  & 10.122  & 8.924  \\
        58870.385 & 10.880  & 10.107  & 8.906  \\
        58871.349 & 10.891  & 10.100  & 8.906  \\
        58884.344 & 10.948  & 10.171  & 9.014  \\
        58891.351 & 10.895  & 10.081  & 8.920  \\
        58908.203 & 10.988  & 10.133  & 8.900  \\
        58909.272 & 10.981  & 10.135  & 8.867  \\
        58911.257 & 10.969  & 10.132  & 8.881  \\
        58914.269 & 10.907  & 10.068  & 8.851  \\
        58919.219 & 10.928  & 10.110  & 8.873  \\
        58920.283 & 10.927  & 10.102  & 8.863  \\
        \hline
    \end{tabular}
\end{table}
\renewcommand{\baselinestretch}{1}
\renewcommand{\baselinestretch}{0.85}
\begin{table*}
    \centering
    \caption{$YJHK$-photometry for IRAS\,07253-2001 in 2021--2023}
    \label{tab:tabl4} \medskip
    \begin{tabular}{c|c|c|c|c||c|c|c|c|c}
        \hline
        JD, 2400000+ & $Y$, mag & $J$, mag & $H$, mag & $K$, mag &JD, 2400000+ & $Y$, mag & $J$, mag & $H$, mag & $K$, mag \\
        \hline
      59541.565& 11.575& 11.011& 10.174 & 8.970 &59656.210 &    --  & 11.020 &10.152 & 8.939\\
      59547.503&  --     & 11.055& 10.220 & 9.014 &59679.203 &  --  & 10.990 &10.148 & 8.960\\
      59552.396&  --     & 11.052& 10.179 & 8.990 &59890.542 &  --  & 10.905 &10.095 & 8.931\\
      59555.467&  --     & 11.056& 10.187 & 8.980 &59892.544 &  --  & 10.916 &10.107 & 8.956\\
      59565.416& 11.617& 11.091& 10.209 & 9.001 &59895.546 &    --  & 10.929 &10.116 & 8.966\\
      59571.483& 11.607& 11.061& 10.176 & 8.952 &59899.598 &    --  & 10.904 &10.098 & 8.934\\
      59584.435& 11.596& 11.034& 10.158 & 8.968 &59912.470 &    --  & 10.933 &10.118 & 8.938\\
      59599.379&  --     & 11.050& 10.197 & 9.013 &59915.440 &  11.425  & 10.941 &10.124 & 8.949\\
      59602.308&  --     & 11.036& 10.190 & 8.982 &59922.458 &  11.407  & 10.907 &10.081 & 8.877\\
      59606.297&  --     &  --   & --     & 8.971 &59930.432 &  --  & 10.913 &10.079 & 8.896\\
      59609.475&  --     & 10.996& 10.186 & 9.015 &59945.519 &  --  & 10.895 &10.087 & 8.886\\
      59611.350& 11.524& 11.011& 10.163 & 8.960 &59953.477 &    11.385  & 10.900 &10.102 & 8.938\\
      59616.329& 11.515& 11.016& 10.163 & 8.948 &59954.384 &    11.391  & 10.903 &10.105 & 8.938\\
      59623.394& 11.561& 11.034& 10.174 & 8.981 &59958.394 &    11.386  & 10.907 &10.092 & 8.923\\
      59625.258& 11.566& 11.040& 10.194 & 8.991 &59962.366 &    11.354  & 10.881 &10.071 & 8.887\\
      59636.285& 11.533& 11.028& 10.180 & 8.955 &59980.331 &    11.357  & 10.873 &10.068 & 8.908\\
      59639.211& 11.548& 11.033& 10.173 & 8.956 &60000.243 &    11.413  & 10.942 &10.119 & 8.930\\
      59641.270& 11.556& 11.039& 10.167 & 8.964 &60012.281 &    11.433  & 10.947 &10.065 & 8.922\\
      59645.267& 11.577& 11.052& 10.186 & 8.978 &        &      &    &   &      \\
        \hline
    \end{tabular}
\end{table*}
\renewcommand{\baselinestretch}{1}

\subsection{ASAS-SN data}

The observational data from the All Sky Automated Survey for
SuperNovae (Kochanek et al., 2017; Shappee et al., 2014) conducted
on robotic telescopes appeared very useful for the study of the
star's photometric behaviour.

About 920 $V$ brightness estimates with an accuracy of
$0\,.\!\!^{\rm m}02$ were obtained by the ASAS-SN project for
IRAS\,07253-2001 from February~14, 2012 till May~25, 2018
(HJD\,=\,2455972.9--2458264.5). The $g$-band observations started
in 2018. In this work we used the $g$ data obtained from
February~17, 2018 till April~19, 2022
(HJD\,=\,2458226.6--2459689.6). A total amount of 580 $g$
brightness estimates with an accuracy of $0\,.\!\!^{\rm m}02$ was
obtained for the interval.

\subsection{Spectroscopic observations}

Spectroscopic observations of IRAS\,07253-2001 were carried out in
2020 and 2023 on the 2.5-m telescope of the CMO SAI MSU with the
new low-resolution Transient Double-beam Spectrograph (TDS)
equipped with holographic gratings (Potanin et al., 2020).  The
detectors in use are Andor Newton 940P cameras with
$512\times2048$ \mbox{E2V CCD42-10} CCDs. A long slit of width of
$1\,.\!\!^{\prime\prime}0$ was selected which provided the best
spectral resolution but at the cost of losing some light if seeing
was worse than $1\,.\!\!^{\prime\prime}0$. The light losses at the
slit may be different for the program and standard stars due to
varying seeing and the accuracy of centering the star in the slit.
Therefore it's impossible to obtain absolute flux-calibrated
spectra with our spectrograph when a $1\,.\!\!^{\prime\prime}0$
slit is used. The spectra covered the range
\mbox{$\lambda$\,3500--7500}. The spectral resolution was 1300 for
the $\lambda$\,3500--5720 region (blue channel) and 2500 for the
$\lambda$\,5720--7500 region (red channel). The log of
observations can be found in Table~5. The moments of spectroscopic
observations are marked in Fig.~4.

The reduction sequence was performed using a number of
self-developed $\tt Python$ scripts. The processing algorithm is
described in Potanin et al. (2020). Although we did not aim to
derive absolute stellar fluxes, and moreover we used
continuum-normalized spectra, nevertheless it was necessary to
observe standard stars to eliminate small-scale features present
in the spectrum due to the transmission inhomogeneities of the
device and the atmospheric absorption bands. The stars from the
list of spectrophotometric standards compiled at the European
Southern
Observatory\footnote{\url{https://www.eso.org/sci/observing/tools/standards/
spectra/stanlis.html}} were used during observations in 2020. In
2023 we used an A0\,V star HIP\,38789 as a spectrophotometric
standard. It is located close to IRAS\,07253-2001, its spectrum
was obtained with a signal-to-noise ratio $S/N \approx 500$ just
after the object.

\setlength{\tabcolsep}{1.8pt}
\begin{table}
\caption{Log of spectroscopic observations}\label{tab:tabl5} \medskip
\begin{tabular}{c|c|r|c|c}
\hline
\multicolumn{1}{c|}{\multirow{2}{*}{Date}} & HJD      & \multicolumn{1}{c|}{$T_{\rm exp}$,}& \multicolumn{1}{c|}{\multirow{2}{*}{SNR}} &\multicolumn{1}{c}{\multirow{2}{*}{Standard}}\\
    &  $2450000+$  & \multicolumn{1}{c|}{s}      &     &\\
\hline
2020/01/18&8867.4& $300\times 3$&130& BD\,+25~4655\\
2020/12/14&9198.5& $400\times 2$&170&  Feige\,66\\
2023/01/06&9951.4& $900\times 3$&190&HIP\,38789\\
2023/01/10&9955.4& $1200\times 3$&340&HIP\,38789\\
\hline \multicolumn{5}{p{8cm}} {\footnotesize{SNR is a
signal-to-noise ratio for the continuum in the resulting spectrum
near $\lambda$\,6000.}}
\end{tabular}
\end{table}

The continuum-normalized spectrum of the standard star was
processed using the \texttt{pySME}
project\footnote{\url{https://pysme-astro.readthedocs.io/en/latest/}
\label{pysme}} (Piskunov and Valenti, 2017; Wehrhahn et al., 2023)
and we managed to deduce stellar parameters (the effective
temperature $T_{\rm eff}=9630\,K$, the surface gravity $\log
g=3.89$, the overall metallicity ${\rm [Me/H]=0.14}$ and the
microturbulence velocity $\xi_t=2.0$ km\,s$^{-1}$) which fitted
the spectral lines with an accuracy better than 1\%. The simulated
spectrum was integrated with the standard transmission curves for
the
filters\footnote{\url{http://svo2.cab.inta-csic.es/theory/fps/},
Generic} and flux-calibrated based on the $V$ photometry taken
from the Simbad database, whereas the $B$ photometry served to get
interstellar extinction by assuming a standard law of interstellar
extinction. The resulting value $A_V=0\,.\!\!^{\rm m}21$ is in
agreement with the absence of diffuse interstellar bands in the
observed spectrum. This approach makes it possible to reconstruct
a pixel-to-pixel transmission curve and to remove telluric
absorptions from the observed spectrum under the condition of
observing at the same airmass and with the full width of the slit
being filled with stellar light. The latter condition is due to
the fact that non-uniform illumination of the slit affects the
line profiles; the difference in the profiles of the tellurics
between the object and the standard will lead to incomplete
compensation of them when divided by the transmission curve and to
the appearance of residual artefacts in the spectrum. A similar
effect will arise if a shift of wavelength occurs between the
observations of the object and the standard due to device
deformations.

\section{OBSERVATIONAL DATA ANALYSIS}

\subsection{Search for periodicity}

A preliminary examination of the photometric data showed that the
star's brightness varies quasi-periodically with time. In order to
determine a period we used dense data sets of the $V$ and $g$
photometry obtained by ASAS-SN (Fig.~2).

To perform a frequency analysis we used the program
\texttt{WINEFK} developed by
V.~P.~Goranskij\footnote{\url{http://www.vgoranskij.net/software/WinEFrusInstruction.pdf}}
which implements a discrete Fourier transform for sets of data
with arbitrary spacing in time (Deeming, 1975).

\begin{figure*}
    \centering
    \includegraphics[scale=0.62] {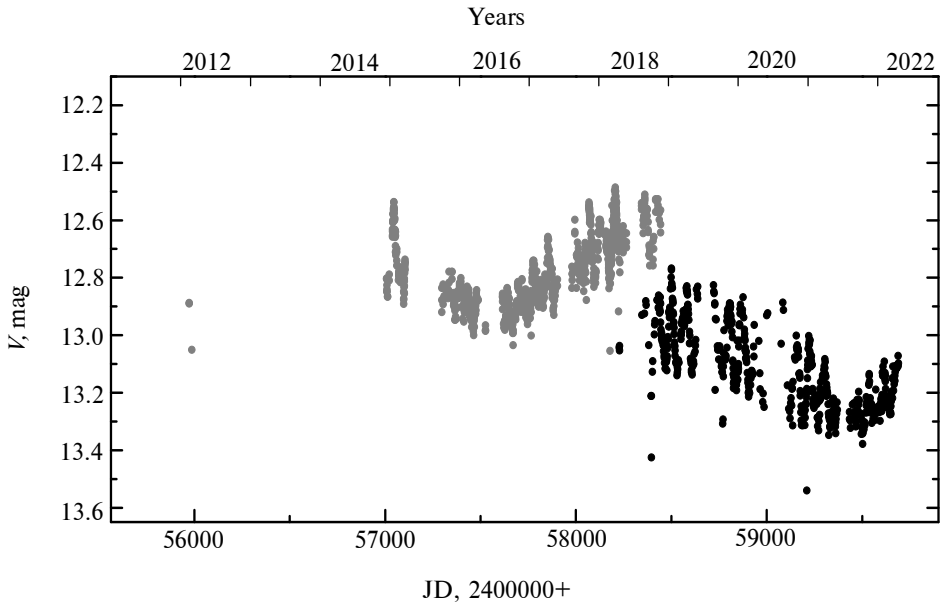}
    \caption{The $V$ (grey dots) and $g$ (black dots) ASAS-SN light curves spanning the intervals 2012--2018 and 2018--2022, respectively.}
    \label{fig2}
\end{figure*}

First, we removed the trend which was satisfactorily fitted with a
quadratic polynomial. Then, we employed the Fourier transform and
successive whitening -- the subtraction of phase-smoothed periodic
variation from the observed light curve. As a result, three
periodic components in the $V$ light curve for 2012--2018 were
identified: the primary period of $P =73\,.\!\!^{\rm d}0$ and,
after successive pre-whitening the data, \mbox{$P = 67\,.\!\!^{\rm
d}8$} и $P =43\,.\!\!^{\rm d}1$. The $g$ data for
\mbox{2018--2022} processed similarly yielded the primary period
of $P = 73\,.\!\!^{\rm d}4$ and, after whitening applied, the
close values of $P = 69\,.\!\!^{\rm d}9$ and $P = 66\,.\!\!^{\rm
d}4$, but also $P = 44\,.\!\!^{\rm d}8$. The amplitude spectra for
the 10--100~days period range obtained from the $V$ and $g$
photometry with the primary periods of $P = 73\,.\!\!^{\rm d}0$
and $P = 73\,.\!\!^{\rm d}4$ marked are shown in Fig.~3 as well as
the phase curves folded on these periods. The maximum peak-to-peak
variations are $\Delta V=0\,.\!\!^{\rm m}3$ and $\Delta
g=0\,.\!\!^{\rm m}35$.

\begin{figure*}
    \includegraphics[scale=1.9]
    {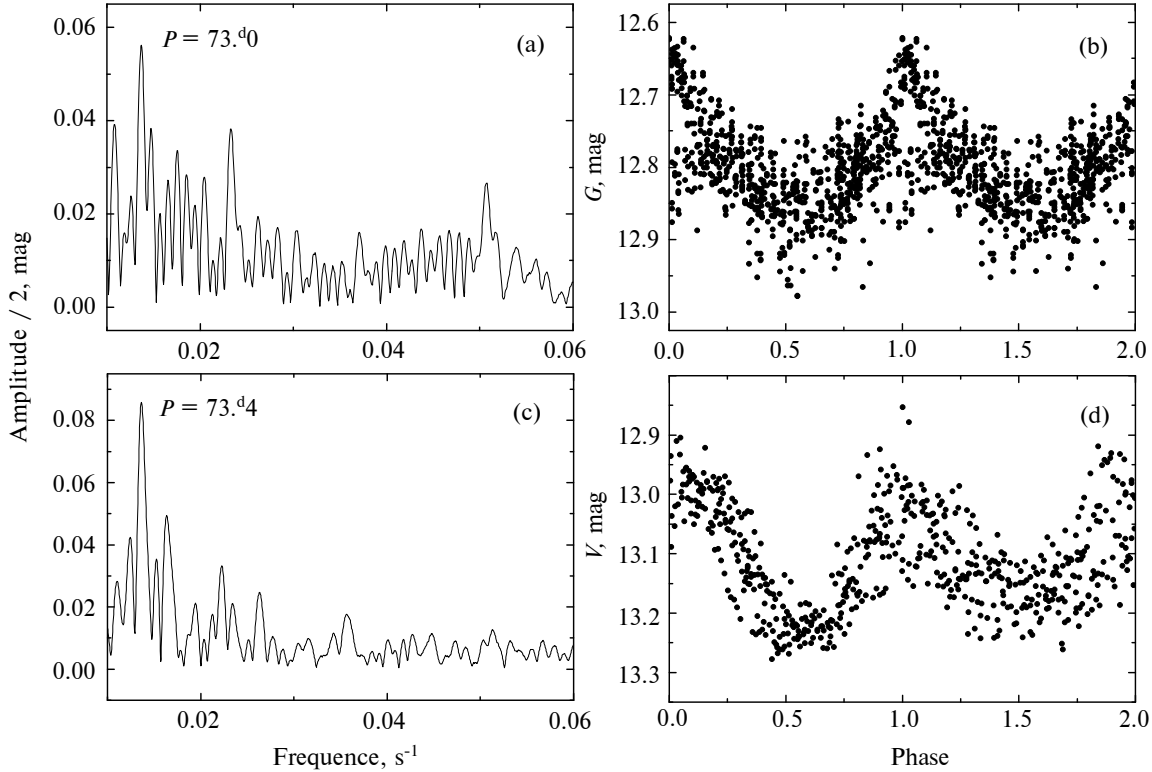}
    \caption{Amplitude spectra for the $V$ (a) and $g$ (c) ASAS-SN data \
    and phase curves folded on the periods corresponding to the most significant
    peaks in the amplitude spectra (b, d).}
    \label{fig3}
\end{figure*}

The detected brightness variability of IRAS\,07253-2001 is typical
for the F0--F8 supergiants at the post-AGB stage of evolution.
Semi-regular brightness variations of these stars are
characterized by small amplitudes (from $0\,.\!\!^{\rm m}1$ to
$0\,.\!\!^{\rm m}6$), the periods of 30--100~days, switching
between modes of close frequencies---the properties which were
described in Sasselov (1984) for the UU\,Her type stars and
confirmed later for a number of other post-AGB objects (Arkhipova
et al., 1993; Hrivnak and Lu, 2000; Kiss et al., 2007).  A
combined study of light, colour and radial velocity curves
supports the idea that these stars vary in brightness due to
pulsations (Hrivnak et al., 2013, 2018).

\subsection{Multicolour photometry analysis}

The $UBVR_CI_C$ light and $U-B$, $B-V$, \mbox{$R_C-I_C$} colour
curves resulted from our observations performed on the RC600
telescope during four seasons of visibility in 2019--2023 are
shown in Fig.~4.

\begin{figure*}
    \includegraphics[scale=1.53]
    {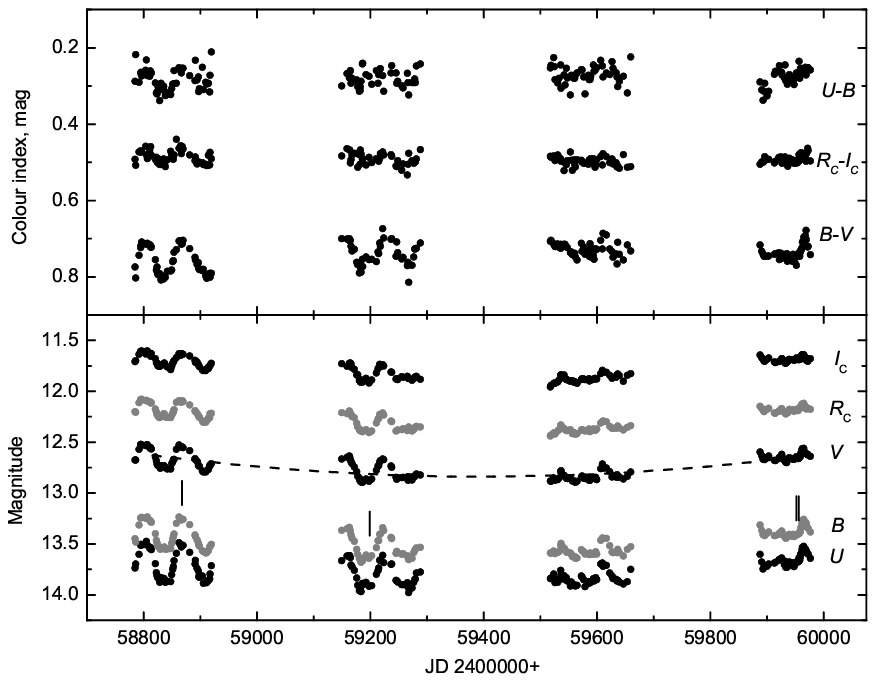}
    \caption{Light and colour curves based on the RC600 data obtained in 2019--2023. The dashed line corresponds to a quadratic polynomial fit of the $V$ data. Vertical line segments indicate the moments of acquiring spectra.}
    \label{fig4}
\end{figure*}

It's clearly seen that the star undergoes semi-regular brightness
oscillations in the $UBVR_CI_C$ bands with varying amplitude
superimposed on a long-term trend. The maximum amplitudes of
brightness variations were observed during the first two seasons:
$\Delta U=0\,.\!\!^{\rm m}40$, $\Delta B=0\,.\!\!^{\rm m}35$,
$\Delta V=0\,.\!\!^{\rm m}25$, $\Delta R_C=0\,.\!\!^{\rm m}22$ and
$\Delta I_C=0\,.\!\!^{\rm m}19$.

With the trend fitted by a quadratic polynomial removed, the
frequency analysis yielded a primary period of $P_0=73\,.\!\!^{\rm
d}3$. Using the \texttt{WINEFK} program we subtracted a
phase-smoothed periodic oscillation with $P_0=73\,.\!\!^{\rm d}3$
from the observed light curve and found a close period
\mbox{$P_1=69\,.\!\!^{\rm d}8$}. The values are in good agreement
with the periods derived from the ASAS-SN data.

We obtained near-IR photometry for six seasons during 2018--2023.
Figures~5 shows the $YJH$ light and $J-H$, $H-K$ colour curves.


\begin{figure*}
    \includegraphics[scale=1.53]
    {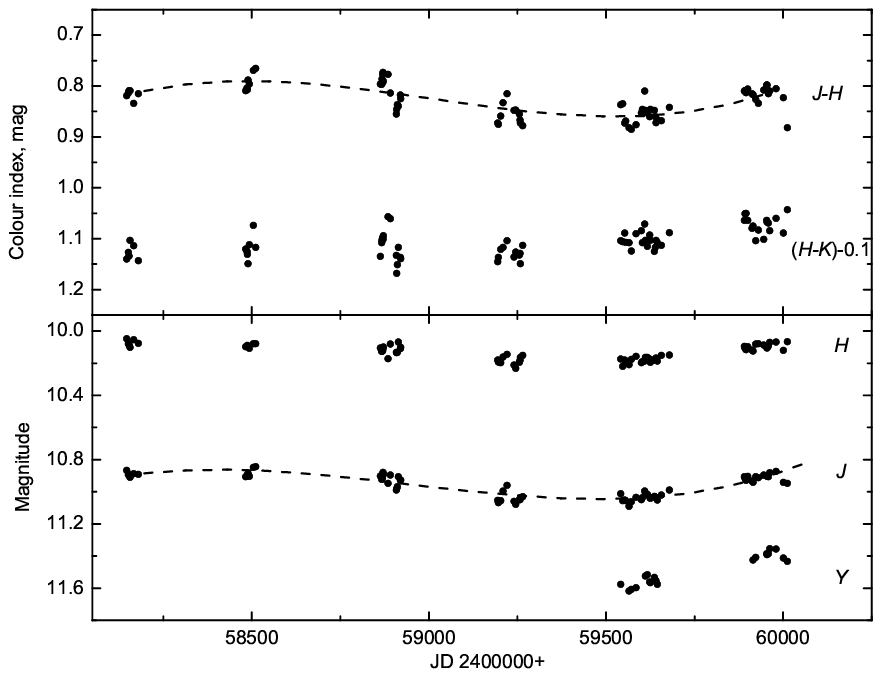}
    \caption{Near-IR light and colour curves spanning 2018--2023. The dashed line corresponds to a cubic polynomial fit of the $J$ data.}
   \label{fig5}
\end{figure*}

The near-IR variations in the $YJHK$ bands are about
$0\,.\!\!^{\rm m}15$ within each season that is larger than the
observational error. The near-IR measurements are less numerous
than the optical ones and the expected amplitude of oscillations
is smaller as well, so we were not able to detect a periodic
component of brightness variations. A long-term trend can be
traced well and is similar to that in the optical.

During the first two seasons of optical observations when the
periodic oscillations were more prominent a clear correlation
between the brightness and the $B-V$ and $R_C-I_C$ colours was
seen (Fig.~6): the star was redder when fainter, which is
indicative of temperature changes due to pulsations. The $U-B$
colour correlation with brightness is less pronounced for the
first season (black dots) and almost absent for the second one
(grey dots). One can also see that the mean brightness in all the
bands was lower in 2020--2021 (grey dots) whereas the season-mean
colours did not change.

\begin{figure*}
   \includegraphics[scale=0.63]
   {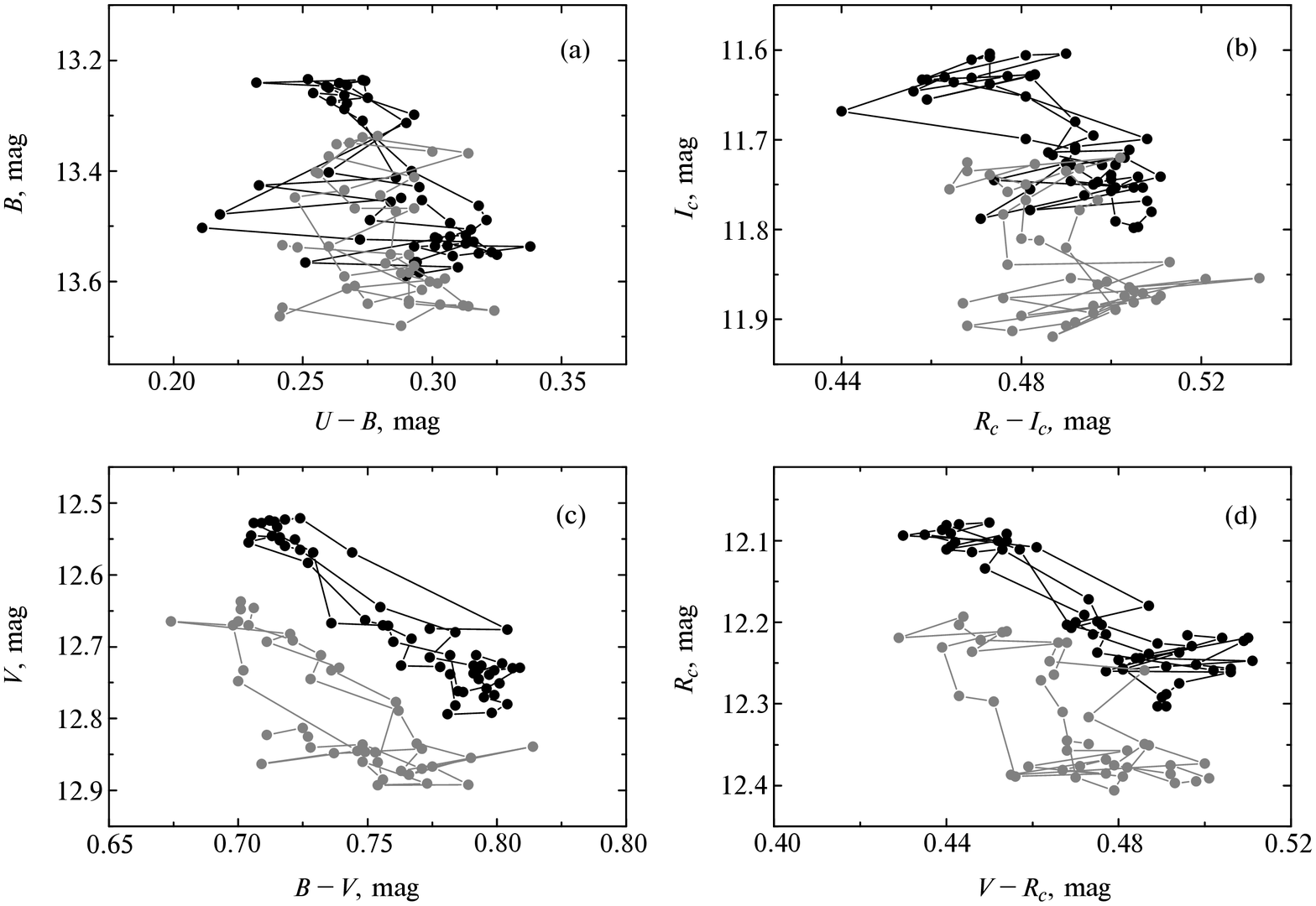}
   \caption{The ``colour--brightness'' diagrams showing the 2019--2020 (black dots) and 2020--2021 (grey dots) data.}
    \label{fig6}
\end{figure*}

\subsection{Long-term trend of brightness}

Figure~7 shows the summary $V$ light curve incorporating the
ASAS-SN and RC600 data over the 2014--2023 interval. Figure~7
implies that there is a sine wave with a large enough period. We
processed the 2014--2023 $V$ data set with \texttt{WINEFK} and in
the 500--3000~days range we found a period of
$P=1810\,\pm\,200$~days. Figures~8 and 9 show the phase $VJHK$
light and $U-B$, $B-V$, $R_C-I_C$, $J-H$, $H-K$ colour curves,
respectively, folded on this period. The optical and near-IR
brightness varies with phase as well as the $J-H$ colour does,
whereas the $U-B$, $B-V$, $R_C-I_C$ and $H-K$ colours do not.

\begin{figure*}
    \includegraphics[scale=0.6]
    {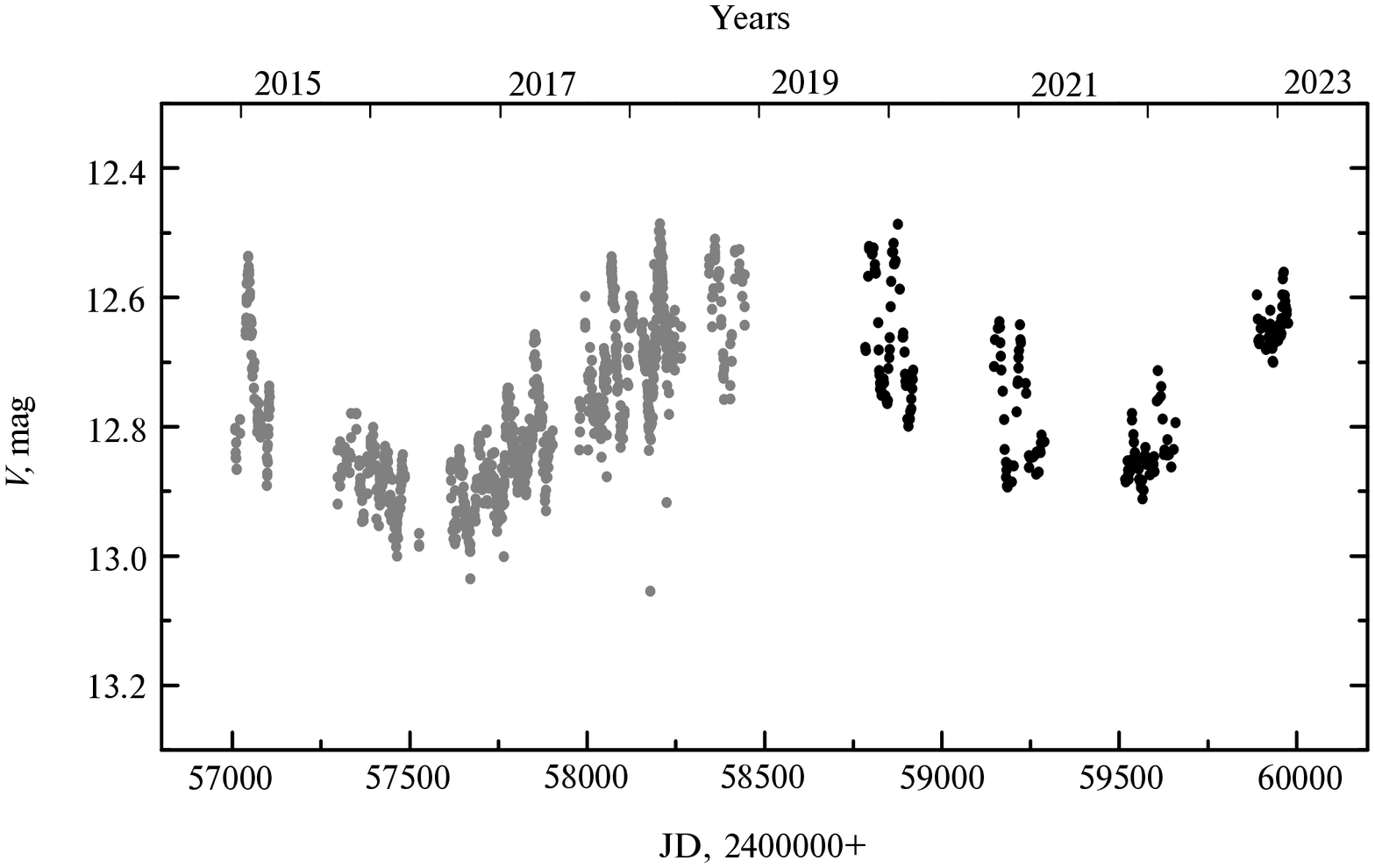}
    \caption{The $V$ light curve based on the ASAS-SN (grey dots) and RC600 (black dots) data.}
    \label{fig7}
\end{figure*}

\begin{figure*}
    \includegraphics[scale=0.65]    {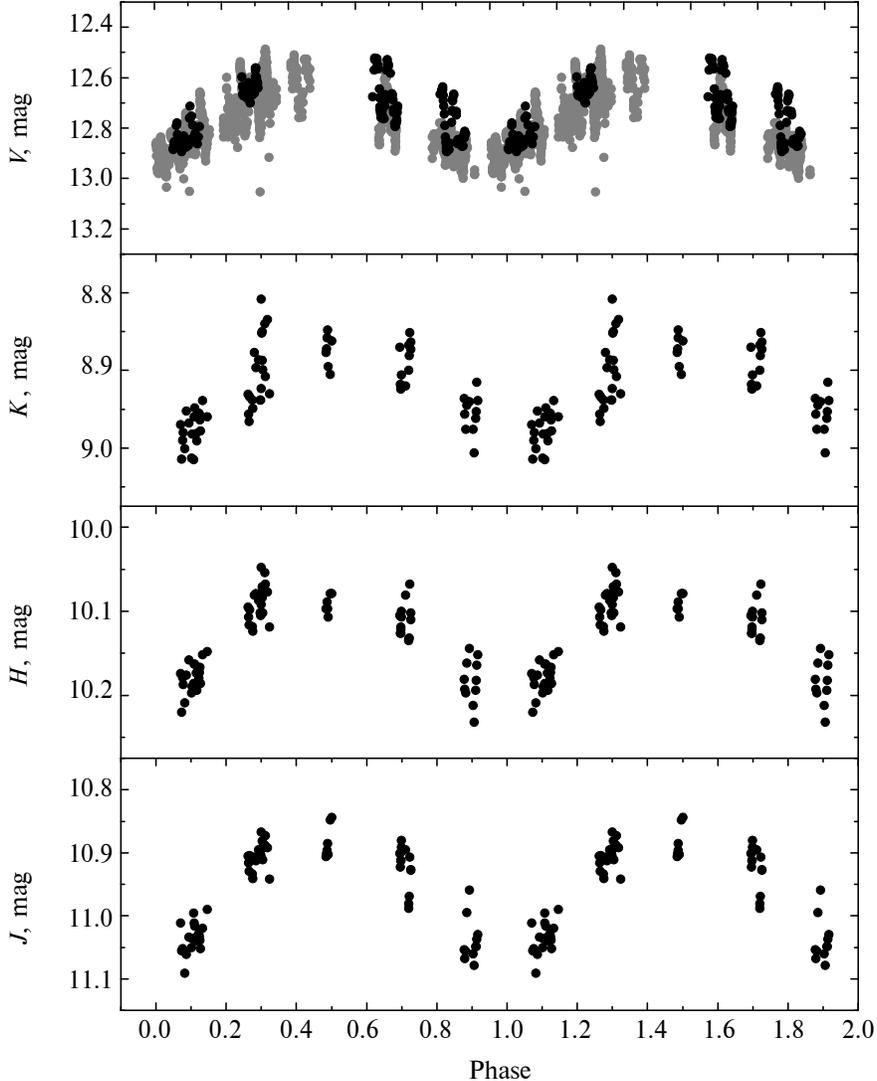}
    \caption{Phase light curves folded on the period of $P=1810^d$ incorporating the ASAS-SN (grey dots) and CMO (black dots) data.}
    \label{fig8}
\end{figure*}

\begin{figure*}
    \includegraphics[scale=0.65]    {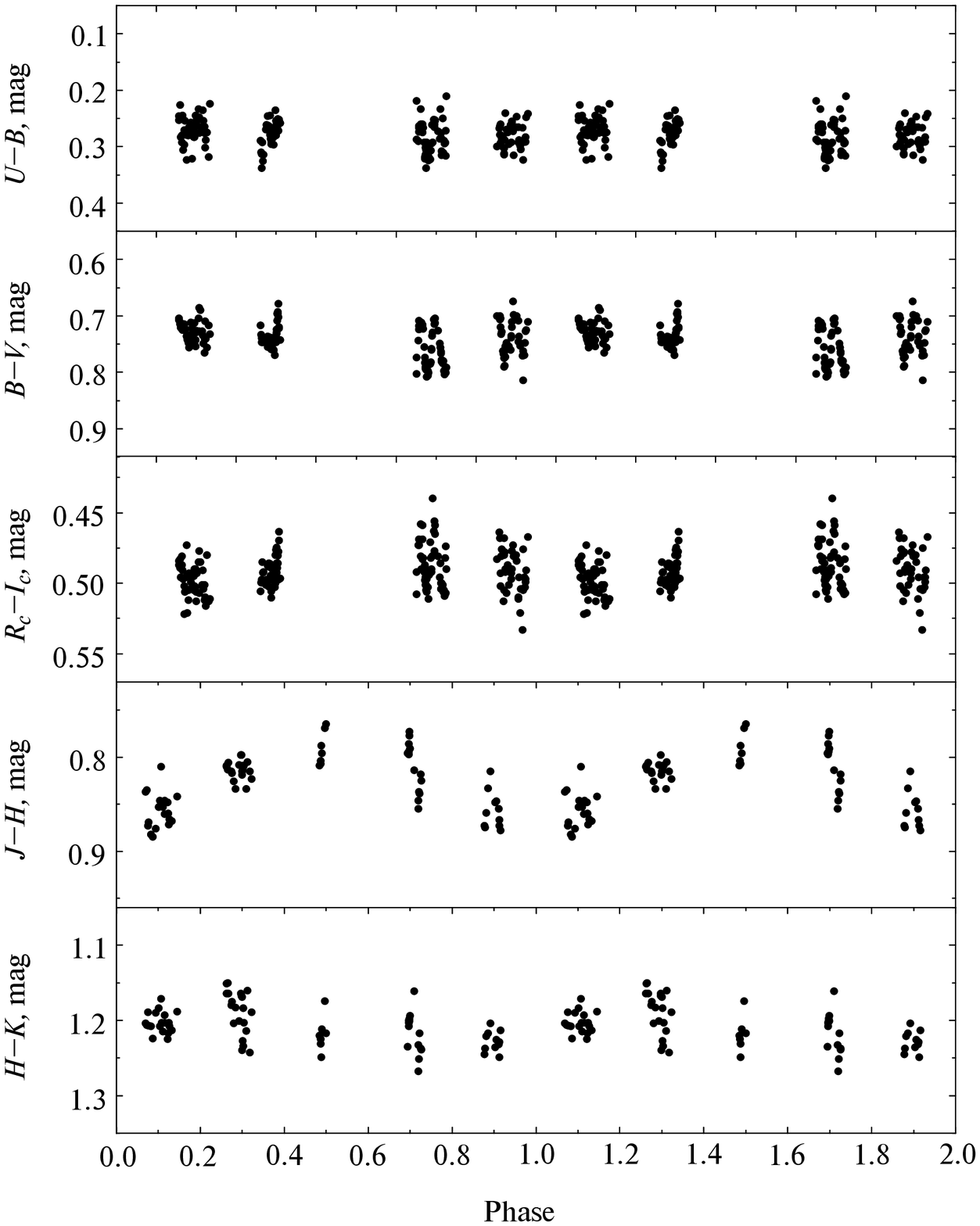}
    \caption{Phase colour curves folded on the period of $P=1810^d$ based on the CMO data.}
    \label{fig9}
\end{figure*}

The long-term brightness variability with a period of about
1800~days which we have found is not surprising for post-AGB
objects. For example, the RV\,Tau stars of the RVb subtype which
are post-AGB stars as well demonstrate a long-term modulation of
mean brightness with the periods of 470--2800~days in addition to
their pulsation activity {Soszy\'{n}ski}  et al., 2017). In our
current understanding, this type of variability is considered
related to binarity and the presence of a circumbinary dusty disc
which produces variable obscuration of the central source due to
orbital motion  (Kiss and B\'{o}di, 2017).

Hotter post-AGB objects, which have already left the instability
strip, sometimes show a long-term modulation of mean brightness,
too. V510\,Pup (IRAS\,08005-2356), a bipolar protoplanetary nebula
with a binary central star, may serve as an example. Manick et al.
(2021) discovered its brightness variability with a period of
$P=2654\,\pm\,124$~days based on optical ($V$) and near-IR
($JHKL$) photometric data. The authors also performed
spectroscopic monitoring of the star and detected a variation of
radial velocity with the same period of $P=2654$~days which they
adopted as the orbital period of the system.

Thus, taking into consideration the fact that some post-AGB
objects demonstrate a long-term periodic trend of brightness we
consider that the found period is orbital and the most likely
reason for this type of brightness variability with optical
colours being constant is the varying obscuration of the central
source by large particles of the dusty disc which produce neutral
absorption during the orbital motion as was detected for
IRAS\,19135+3937 (Gorlova et al. (2015) and our unpublished data).

\subsection{Analysis of spectroscopic data}

In total, we obtained four spectra for IRAS\,07253-2001 in 2020
and 2023. The best-quality spectrum (with the highest SNR) was
acquired on January~10, 2023 under very good weather conditions
with the longest exposures. That spectrum was mainly used for the
following analysis. Figure~10 shows the continuum-normalized
spectrum for that date with the identified lines marked. For line
identification we used the VALD3 database (Ryabchikova et al.,
2015).

\begin{figure*}
    \centering
    \includegraphics[scale=0.635]
    {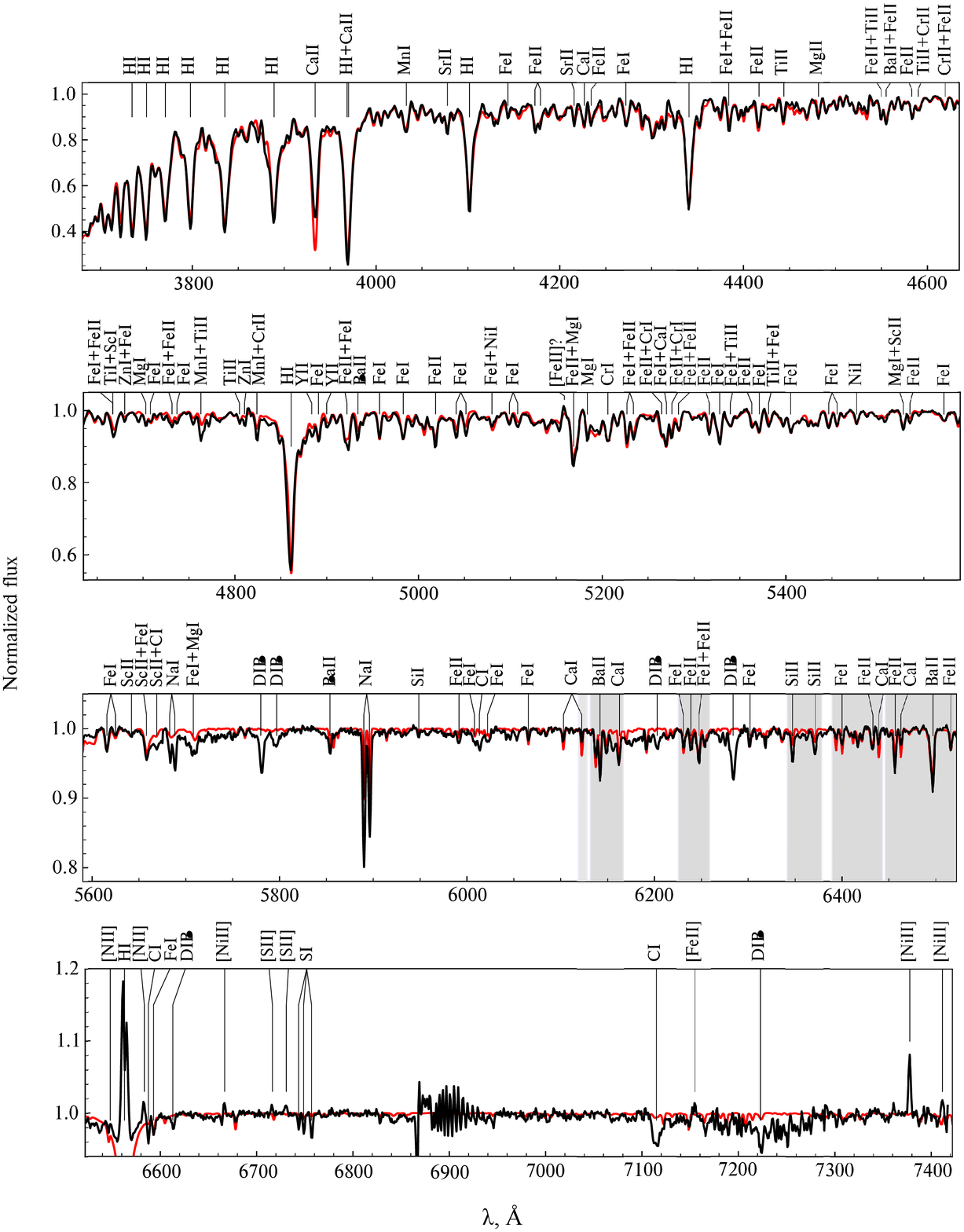}
    \caption{Continuum-normalized spectrum obtained on January~10, 2023 with
    the identified lines marked. Grey regions were used for radial velocity measurements.
    The red line corresponds to the model spectrum (in electronic form of the paper).}
    \label{fig10}
\end{figure*}

In addition to hydrogen lines there are numerous absorptions of
neutral and once-ionized metals Fe\,I, Fe\,II, Mg\,I, Mn\,I,
Sc\,II, Ni\,I, Si\,II, etc. The presence of strong lines of S\,I
($\lambda$\,6744, $\lambda$\,6749, $\lambda$\,6757) and C\,I
($\lambda$\,6010--6020, $\lambda$\,6588,
\mbox{$\lambda$\,7107--7120)} is worth mentioning. We also
detected the lines of $s$-process elements: barium Ba\,II
($\lambda$\,5853, $\lambda$\,6142, $\lambda$\,6498), strontium
Sr\,II ($\lambda$\,4078 and $\lambda$\,4215) and yttrium Y\,II
($\lambda$\,4884). Broad absorptions at $\lambda$\,5797,
$\lambda$\,6286, $\lambda$\,6613 may be identified as diffuse
interstellar bands (DIBs). In the spectrum of IRAS\,07253-2001 the
K\,Ca\,II line ($\lambda$\,3933) is shallower than the blend
H$\epsilon$+H\,Ca\,II in contrast to what is seen in the spectra
of other post-AGB stars of close spectral classes (V887\,Her,
V1648\,Agl and V448\,Lac) where these lines are quite equal in
depth  (Hrivnak et al. (1989); {Su\'{a}rez} et al. (2006) and our
unpublished data).

\subsubsection{Envelope emission lines}

An important feature of the spectrum is the presence of the
emission component of H$\alpha$. The H$\alpha$ profile is shown in
Fig.~11 where we plot the spectra obtained on January~18, 2020,
December~14, 2020, and January~10, 2023. We do not show the
spectra obtained on January~6, 2023 as there is almost no
difference in the H$\alpha$ profile with that from January~10,
2023 (Fig.~11).

\begin{figure}
    \includegraphics[scale=1.3]
    {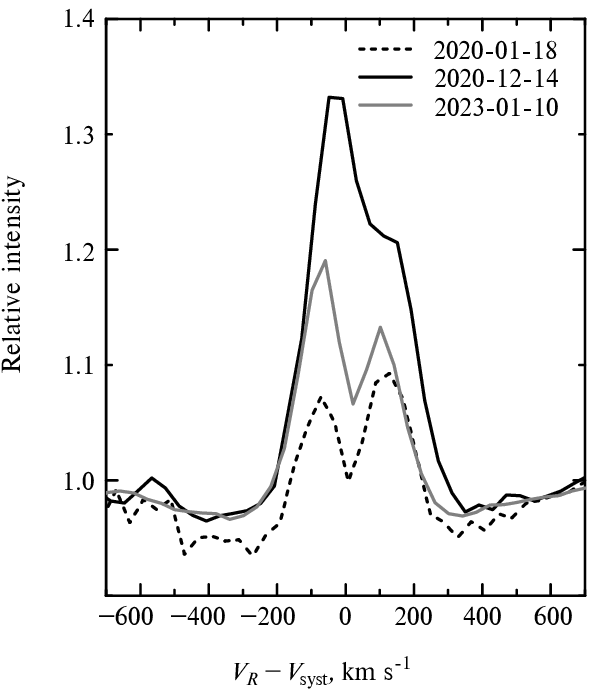}
    \caption{The H$\alpha$ profiles in the normalized spectra obtained on January 18, 2020, December 14, 2020, and January 10, 2023.}
    \label{fig11}
\end{figure}

The H$\alpha$ line demonstrates a double-peaked emission component
and varies significantly with time. The times of spectroscopic
observations are marked in Fig.~4. On January~18, 2020 when the
star was at pulsation maximum with $V=12\,.\!\!^{\rm m}53$, the
emission component was the faintest. The next spectroscopic
observation carried out on December~14, 2020 coincided with a
pulsation minimum when the brightness was $V=12\,.\!\!^{\rm m}89$.
At that moment the H$\alpha$ emission in the normalized spectrum
appeared considerably stronger, with the central absorption almost
absent. On January~10, 2023 at $V=12\,.\!\!^{\rm m}66$ the
emission component was double-peaked again and had intermediate
normalized intensity.

The emission equivalent width variation can be explained if we
assume that the stellar continuum varies in brightness whereas the
emission comes from a circumbinary gaseous envelope and has
constant intensity. Thus, a double-peaked profile arises when the
emission is added on the photospheric absorption.

We did not expect to detect any forbidden lines in the spectrum of
a cool star and surprisingly we have found the emissions [N\,II]
$\lambda$\,6548, $\lambda$\,6584, [S\,II] $\lambda$\,6716,
$\lambda$\,6731, [Ni\,II] $\lambda$\,6667, $\lambda$\,7378,
$\lambda$\,7412 and [Fe\,II] $\lambda$\,7155. As far as we know,
this phenomenon has not been observed for F0--F8 post-AGB
supergiants unlike hot post-AGB stars with $T_{\rm eff} >
15\,000$~K which spectra are a sum of radiation from the central
star and a low-excitation gas shell as it was shown for
IRAS\,14331-6435  (Arkhipova et al., 2018) and IRAS\,18379-1707
(Ikonnikova et al., 2020).

We list the equivalent widths for [N\,II] $\lambda$\,6584, [S\,II]
$\lambda$\,6716, $\lambda$\,6731, [Ni\,II] $\lambda$\,6667,
$\lambda$\,7378, $\lambda$\,7412 in Table~6. The lines being
faint, the uncertainties in the measurements of equivalent widths
are as large as 15--20\%.

\begin{table}[h!]
\caption{Equivalent widths of emission lines in the spectrum of IRAS\,07253-2001 for January~10, 2023}\label{tab:tabl6} \medskip
\begin{tabular}{l|c|c}
\hline
\multicolumn{1}{c|}{Line}&$\lambda_{\rm lab}$, \AA &$EW$, \AA\\

\hline
[N\,II]\,1F & 6583.45&0.057\\

[Ni\,II]\,2F& 6666.80&0.037\\

[S\,II]\,2F & 6716.47&0.024\\

[S\,II]\,2F & 6730.85&0.036\\

[Ni\,II]\,2F& 7377.83&0.221\\

[Ni\,II]\,2F& 7411.61&0.108\\

\hline
\end{tabular}
\end{table}

The line intensity ratios [S\,II]
$F(\lambda\,6716)/F(\lambda\,6731)$ and [Ni\,II] $F(\lambda\,
6667)/F(\lambda\,7378)$ barely depend on electron temperature
$T_e$ and may be used to estimate the electron density $N_e$ in
the region where the emissions arise. We have rejected the
[Ni\,II] $\lambda\,7412$ line from the analysis as it is located
at the edge of our spectral range and is possibly distorted.

Before we compare the observed intensity ratios with the predicted
ones, it's necessary to correct them for reddening. Reddy and
Parthasarathy (1996)  give the value $A_V=0\,.\!\!^{\rm m}90$ (or
$E(B-V)=0\,.\!\!^{\rm m}29$) for the interstellar extinction and
$A_V=2\,.\!\!^{\rm m}1$ (or $E(B-V)=0\,.\!\!^{\rm m}68$) for the
total one which includes the circumstellar part. Vickers et al.
(2015) adopted \mbox{$E(B-V)=0\,.\!\!^{\rm
m}46\,\pm\,0\,.\!\!^{\rm m}05$}. As we don't know the distance to
the object (see Section 4), we can't yet estimate $E(B-V)$ using
the interstellar extinction maps.

Taking into account the ambiguity of $E(B-V)$, we derived the
intensity ratios for [S\,II]
$F(\lambda\,6716)/F(\lambda\,6731)\approx0.67$ and [Ni\,II]
$F(\lambda\,6667)/F(\lambda\,7378)\approx0.20$ using the
equivalent widths for these lines and the modelled spectral energy
distribution (see~Section~3.4.2 for details). To estimate $N_e$
for the region where the [S\,II] and [Ni\,II] emissions originate
we compared the observed ratios with the theoretical ones based on
the emissivities calculated under non-LTE conditions by Giannini
et al. (2015). We derived $N_e=(1.5$--$2.5)\times 10^3$~cm$^{-3}$
for the [S\,II] zone and a significantly larger value of
\mbox{$N_e=(1$--$3)\times10^6$~cm$^{-3}$} for the [Ni\,II] region
for the temperature range $T_e=5000$--$15\,000$~K. This result is
in agreement with the conclusion of Bautista et al. (1996)
articulated for gas nebulae that the value of $N_e$ derived from
the [Ni\,II] lines appears larger than that derived from [S\,II].

The origin of forbidden emission lines seen in the spectrum of
IRAS\,07253-2001 is still questionable. If we consider
IRAS\,07253-2001 binary then the forbidden lines may point to the
presence of a hot component in the system, most likely a white
dwarf, and to detect it ultraviolet observations would be needed.

\subsubsection{Determination of the stellar parameters}

We have tried to determine the stellar parameters by comparing the
observed spectrum with the synthetic one found by the pySME
program (see the reference above) using the MARCS model
atmospheres (Gustafsson at el., 2008).

A model with $T_{\rm eff}=6300 \pm 300$~К, \mbox{$\log g=2.0\pm
0.6$}, $\xi_t=4.0 \pm 1.7$~km\,s$^{-1}$, [Me/H]\,$=-1.2 \pm 0.2$
fits the hydrogen lines and most of the metal lines well with an
exception for the Ca\,II lines which appear stronger in the model
spectrum than in the observed one and for the S\,I and C\,I lines
enhanced in our spectrum but absent in the model one (Fig.~10). To
obtain more reliable stellar parameters and to estimate abundances
in the atmosphere a high-resolution spectroscopy combined with a
non-LTE approach would be needed.

This star with its lower metallicity along with the presence of
strong enough lines of S and C provides a good illustration of the
peculiar abundances of some post-AGB stars noted in early works,
e.g. Waelkens et al. (1991) for HD\,52961. Thus, some A--F
supergiants demonstrate almost solar abundances of C, N, O, S and
Zn but the abundances of Fe, Mg, Ca, Si, Cr and some other
elements are much lower than solar. Lamers (1992) suggested that
the atmospheres of post-AGB stars were originally solar in
composition but at the end of the AGB stage some heavy nuclei were
captured by dust grains and blown away from the atmosphere. So,
the present atmospheres of these stars are comprised of
heavy-elements-depleted gas.

Now we are going to compare the derived parameters with the
results published previously.

Using a low-resolution spectrum covering $\lambda$\,5800--8500,
Reddy and Parthasarathy (1996) derived an F5\,I(e) spectral class
for the star. An effective temperature of $T_{\rm eff} = 7000$~K
was determined based on the $T_{\rm eff}$--spectral class
calibration proposed by Flower (1977).  Note that the
{Strai\v{z}ys} (1982) calibration gives $T_{\rm eff} = 6500$~K for
this spectral and luminosity class. Reddy and Parthasarathy (1996)
adopted the value of $\log g=1.0$ from the tables with the $\log
g$--luminosity relations (Flower, 1977).

IRAS\,07253-2001 was assigned an F2\,I spectral type by
{Su\'{a}rez}  et al. (2006) based on a low-resolution spectrum
(spectral dispersion of 2.47\,\AA\,pixel$^{-1}$) covering
\mbox{$\lambda$\,4272--6812}. Molina (2018)  derived $T_{\rm eff}
= 7826 \pm 91$~K relying on the same spectrum and the empirical
relation \mbox{$T_{\rm eff} = (8114 \pm 65) + (146 \pm 24)
(\rm{Ca\, II\,K})$} which incorporates the equivalent width of
Ca\,II\,K ($\lambda\,3933$). Note that the line used was out of
the observed wavelength coverage. Other parameters presented by
Molina (2018), namely [Fe/H]\,$=-0.81 \pm 019$ and $\log g = 1.28
\pm 0.21$, do not seem reliable as well, since they were estimated
from the equations given therein and the equations include the
equivalent widths of the Fe\,I ($\lambda\,4271$) blend and Fe,
Ti\,II \mbox{($\lambda\,4172$--$4179$)} lines but the measurements
of these spectral features are not listed in Table~1 of Molina
(2018).

\subsubsection{Radial velocity measurements}

Since the model spectrum fits the most lines quite well, we can
use it for measuring radial velocity. For this purpose we have
selected eight spectral regions with best coincidence. All the
selected areas relate to the red channel because the wavelength
calibration of TDS is much more accurate for the red channel than
for the blue one. Besides, a large number of sky emission lines
allows the correction of calibration shift which arises due to
deformation of the spectrograph. The resulting calibration
accuracy relative to sky lines is about 3~km\,s$^{-1}$. After the
wavelengths are calibrated relative to sky lines, they are
corrected to the solar system barycenter. In order to compute
radial velocity relative to the model spectrum, the latter is
convolved with a Gauss function to fit the widths of absorption
lines. The observed spectrum in each of the selected areas is
additionally normalized and scaled to fit best the continuum level
and line depths of the model spectrum. Radial velocity and two
scaling parameters are computed using the least-square method.

A mean radial velocity averaged over the selected areas was found
to be $V_R=23 \pm 6$~km\,s$^{-1}$ for January~18, 2020, $V_R=6 \pm
6$~km\,s$^{-1}$ for December~14, 2020, $V_R=39 \pm 5$~km\,s$^{-1}$
for January~6 and 10, 2023, taking into account the calibration
error. The difference in radial velocity between December~2020 and
January~2023 is clearly seen at first sight: the absorption lines
are shifted, whereas the interstellar and forbidden lines stay in
place. So, the star shows a variation of radial velocity with an
amplitude not less than 30~km\,s$^{-1}$. A typical range of radial
velocity due to pulsational motions in the atmosphere of post-AGB
stars is about 10~km\,s$^{-1}$ (Hrivnak et al., 2018).  Therefore,
the derived variation of radial velocity may reveal orbital
motion, and moreover the semi-amplitude value of $K_1 \approx
15$~km\,s$^{-1}$ falls into the range of $K_1$ for the known
binary post-AGB stars (see Oomen et al. (2018)  and reference
therein).

\section{CONCLUSION}\label{results}

The primary results of this work are the following:

For the first time a long-term multicolour photometry in the
optical and near-IR range was acquired for the post-AGB object
\mbox{IRAS\,07253-2001}.

Based on the ASAS-SN and our data, low-amplitude quasi-periodic
brightness variability with the main period of about 73~days and
close periods of 68 and 70~days was found. The variation of
brightness amplitude is caused by the beating of close
frequencies. The colour--brightness relation shows evidence for
temperature variations due to pulsations. The variability pattern
and pulsational periods are in agreement with what is observed for
typical post-AGB stars with F0--F8 spectral types  (Arkhipova et
al., 2010, 2011; Hrivnak et al., 2010, 2022).

Long-term brightness variability with a period of about 1800~days
was found based on the ASAS-SN and our multicolour data and this
period is likely orbital.

Based on spectroscopic data of higher resolution that previous
measurements we identified spectral lines and compiled a spectral
atlas. We fitted the spectrum and found the atmospheric parameters
for the star: \mbox{$T_{\rm eff}=6300 \pm 300$~K}, $\log g=2.0 \pm
0.6$, $\xi_t=4.0 \pm 1.7$~km\,s$^{-1}$, [Me/H]\,$=-1.2 \pm 0.2$.

A variation of radial velocity with an amplitude of about
30~km\,s$^{-1}$ was detected which we interpret as an evidence for
binarity.

The H$\alpha$ emission component was shown to be variable. We draw
a conclusion that it originates from the stellar envelope.

Forbidden emission lines radiated by a gas envelope were detected
in the spectrum. We suppose that they are excited by the hot star
in the binary.

The acquired data covering a wide wavelength range from
0.35~$\mu$m (the $U$-band) to 2.2~$\mu$m (the $K$-band) can be
used later for modelling the spectral energy distribution of the
star and determining the dust shell parameters.

There is no doubt about the star's evolutionary status: it is
surely a post-AGB supergiant. But we are not able to assess its
mass by comparing its parameters (effective temperature and
luminosity) with model simulations (e.g., Miller Bertolami
(2016)). We hoped to constrain the luminosity using the GAIA data.
But the parallax is negative in the GAIA\,DR2 catalogue $\pi=-2.2
\pm 0.4$~mas  (Brown et al., 2018). The value $\pi=2.4 \pm
0.4$~mas from the GAIA\,DR3 catalogue (Brown et al., 2021)  yields
a distance of $d=452_{-90}^{+167}$~pc (Bailer-Jones et al., 2021)
and luminosity of $L=25.6_{-9.2}^{+22.4}L_{\odot}$ (Oudmaijer et
al., 2022) which is out of range of the luminosities predicted for
post-AGB models $3000\,L_{\odot}<L<15\,000\,L_{\odot}$ (Miller
Bertolami, 2016). It's worth mentioning that the parameter RUWE
(Renormalised Unit Weight Error) which parametrizes the quality of
astrometric solution is $42.89 \gg 1$ for IRAS\,07253-2001 which
indicates an extremely high uncertainty in parallax that makes it
very unreliable.

\section*{ACKNOWLEDGEMENTS}

This work has been supported by the M.~V.~Lomonosov Moscow State
University Program of Development (scientific and educational
school ``Fundamental and applied space research''). We are
grateful to the staff of the 2.5-m telescope of the CMO SAI MSU
who carried out single observations, namely B.~S.~Safonov,
O.~V.~Vozyakova, O.~V.~Egorov, V.~S.~Lander.

S.~G.~Zheltoukhov and A.~M.~Tatarnikov acknowledge support from
the Russian Scientific Foundation (grant No.~23-22-00182).

\section*{CONFLICT OF INTEREST}

The authors declare that there is no conflict of interest.

\end{document}